\documentclass{appolb}
\usepackage{graphicx, color, subfigure, slashed, listings, fancyhdr, amsmath, amsthm, amssymb, bbm}

\begin{document}
\title{On the Landau gauge matter-gluon vertex in scalar QCD in a functional
approach\thanks{Presented at Excited QCD 2013, 
$3^{rd}$ - $9^{th}$ February 2013, Bjelasnica Mountain,  Sarajevo, Bosnia-Herzegovina}
}
\author{Markus Hopfer and Reinhard Alkofer
\address{Institut f\"ur Physik, Karl-Franzens Universit\"at Graz\\ Universit\"atsplatz 5, 8010 Graz, Austria}
}
\maketitle
\begin{abstract}
Recently the quark-gluon vertex has been investigated in Landau gauge using  a combined Dyson-Schwinger and nPI
effective action approach. We present here a numerical analysis of a simpler system where the quarks  have been
replaced by charged scalar fields.  We solve the coupled system of Dyson-Schwinger equations for the scalar 
propagator, the scalar-gluon vertex and the Yang-Mills propagators in a  truncation related to earlier studies. 
The calculations have been performed for scalars both in the fundamental  and the adjoint representation. A clear
suppression of the Abelian diagram is found in both cases. Thus, within the used truncation the suppression of the
Abelian diagram  predominantly happens dynamically and is to a high degree independent of the colour structure.
The numerical techniques developed here can directly be applied to  the fermionic case.
\end{abstract}
\section{Motivation}

Confinement and dynamical chiral symmetry breaking (D$\chi$SB) are prominent  features of QCD but still not
satisfactorily understood. There is evidence that the quark-gluon vertex  may provide a key position in the
pursuit of finding mechanisms behind these phenomena, see, {\it e.g.}, Ref.\ \cite{Alkofer:2008tt} and references
therein.  Since one is interested in the non-per\-tur\-ba\-tive behaviour of this Green function, appropriate
methods have to be implemented.
Dyson-Schwinger equations (DSEs), suitably combined with other functional methods as, {\it e.g.}, Functional
Renormalization Group Equations or nPI Effective Actions and cross-checked with corresponding lattice results
where available, provide a tool to investigate these low-energy phenomena,
see, {\it e.g.}, Ref.\ \cite{Alkofer:2000wg} and references therein.

It turns out that the quark-gluon vertex is a quite complex multi-tensor object due to its Dirac structure.
Only very recently attempts towards a full self-consistent solution including all possible tensor structures 
have been made \cite{Hopfer:2013np}, where, however, up to now only the non-Abelian  diagram has been taken 
into account. 
However, despite the indications that the Abelian counterpart might be subleading,
only a full self-consistent treatment can prove this assumption. 
Herein we dwell on another possibility to shed light on this issue.
One can substitute the quarks by charged  scalar fields residing in the fundamental representation of 
$SU(3)$. Details and applications of scalar QCD within this context can be found, {\it e.g.}, in 
Refs.\ \cite{Fister:2010yw,Maas:2011yx} as well as references therein.
Thus, one can try to mimic certain aspects of QCD while on the other hand can work with a much 
simpler system due to the absence of Dirac structure. 
We will follow this strategy throughout the following presentation.

\subsection{Scalar QCD}
\label{Sec:scalarQCD}
\begin{figure}[h]
\centerline{%
\includegraphics[width=84mm]{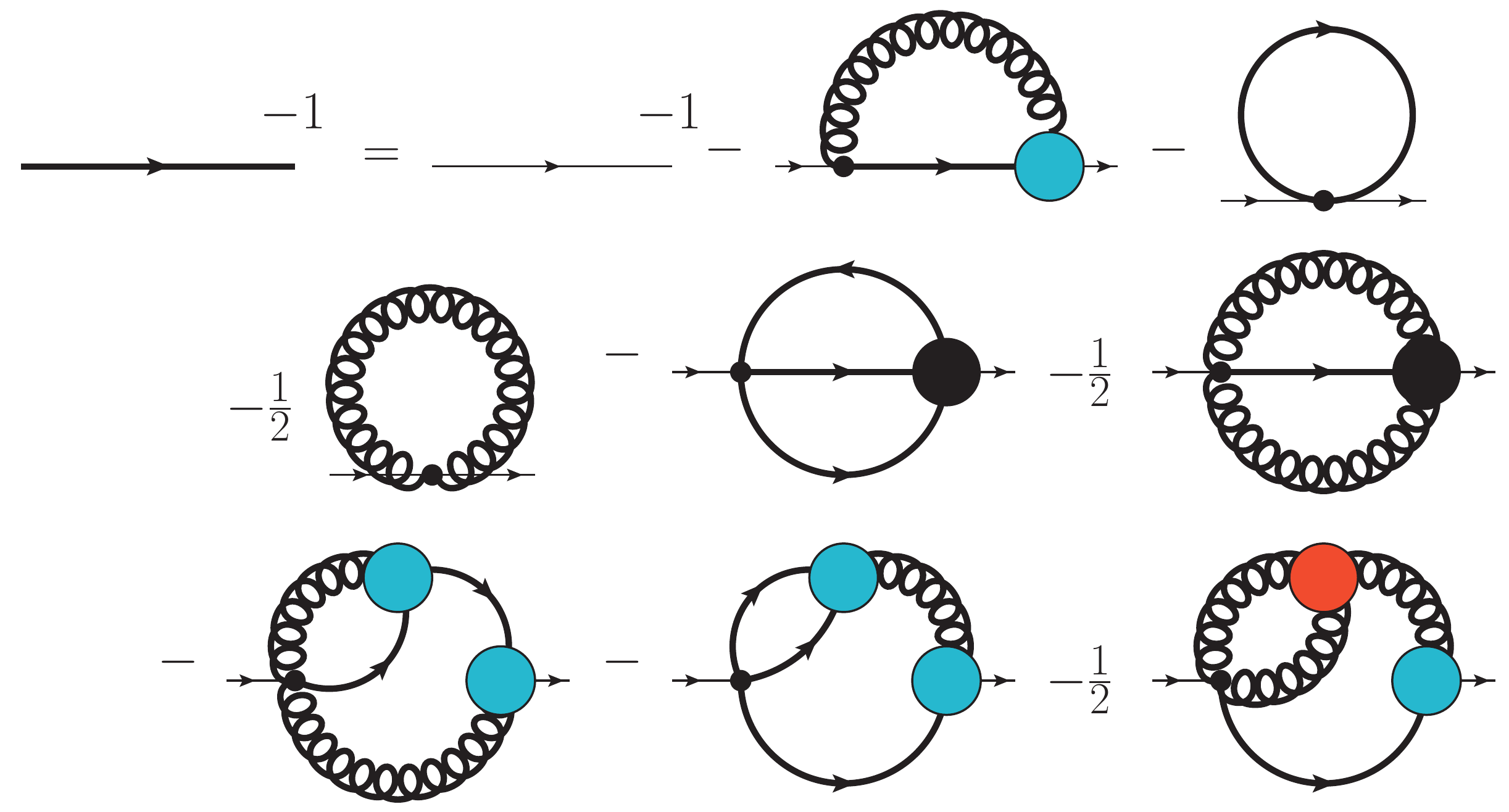}}
\caption{The DSE for the scalar propagator \cite{Fister:2010yw}. 
Full blobs denote dressed vertices, all internal propagators
are dressed.}
\label{Fig:scalar_prop}
\end{figure}

\begin{figure}[h]
\center
\includegraphics[width=69mm]{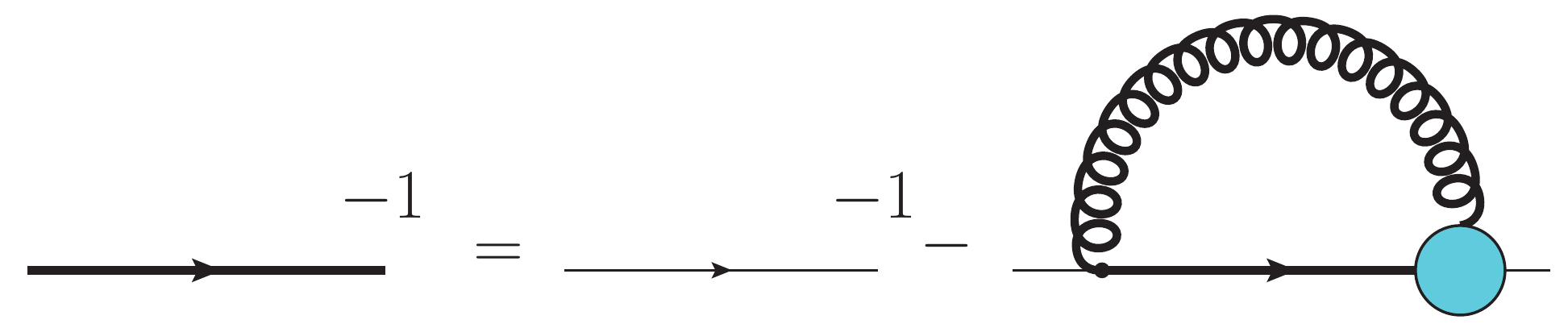}\\
\includegraphics[width=89mm]{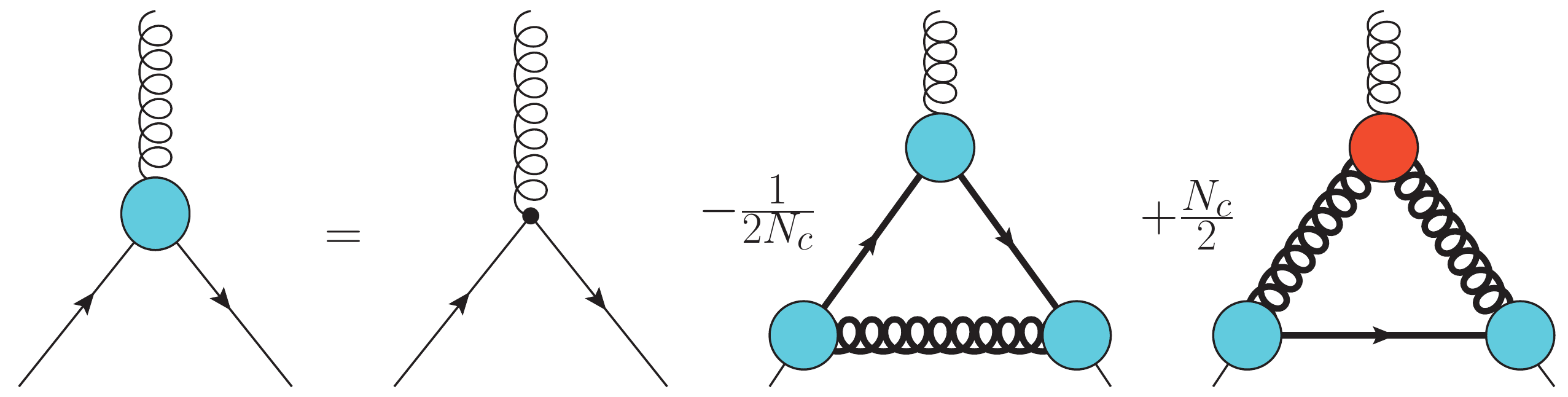}
\caption{The scalar propagator and the scalar-gluon vertex DSE 
in a diagrammatically equivalent truncation to the system treated in
Ref. \cite{Hopfer:2013np}, {\it cf.}, also Ref.\ \cite{Alkofer:2008tt}. All internal propagators are dressed. 
Full blobs denote
dressed vertices. The three-gluon vertex has been modelled according to Ref.\ \cite{Fischer:2003rp}.}
\label{Fig:CoupledSystemScalar}
\end{figure}

\begin{figure}[hb]
\center
~\\[-6mm]
\includegraphics[width=74mm]{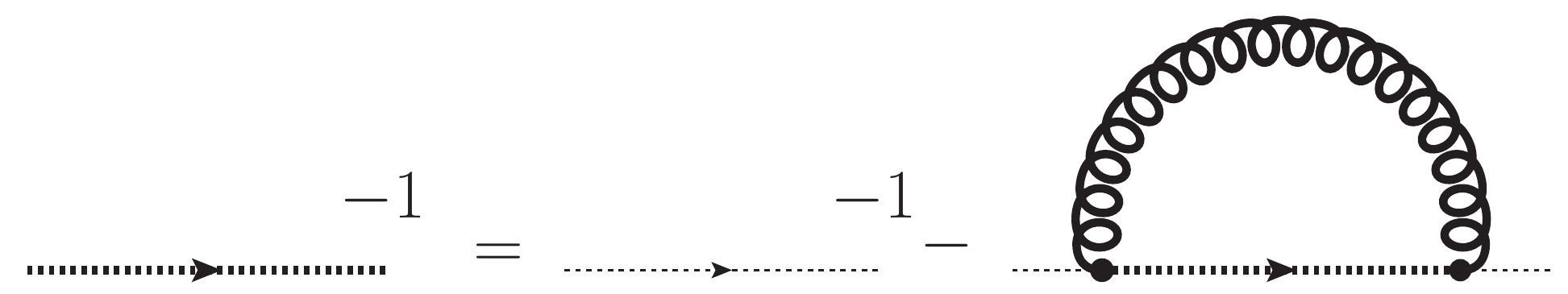}\\[4mm]
\includegraphics[width=74mm]{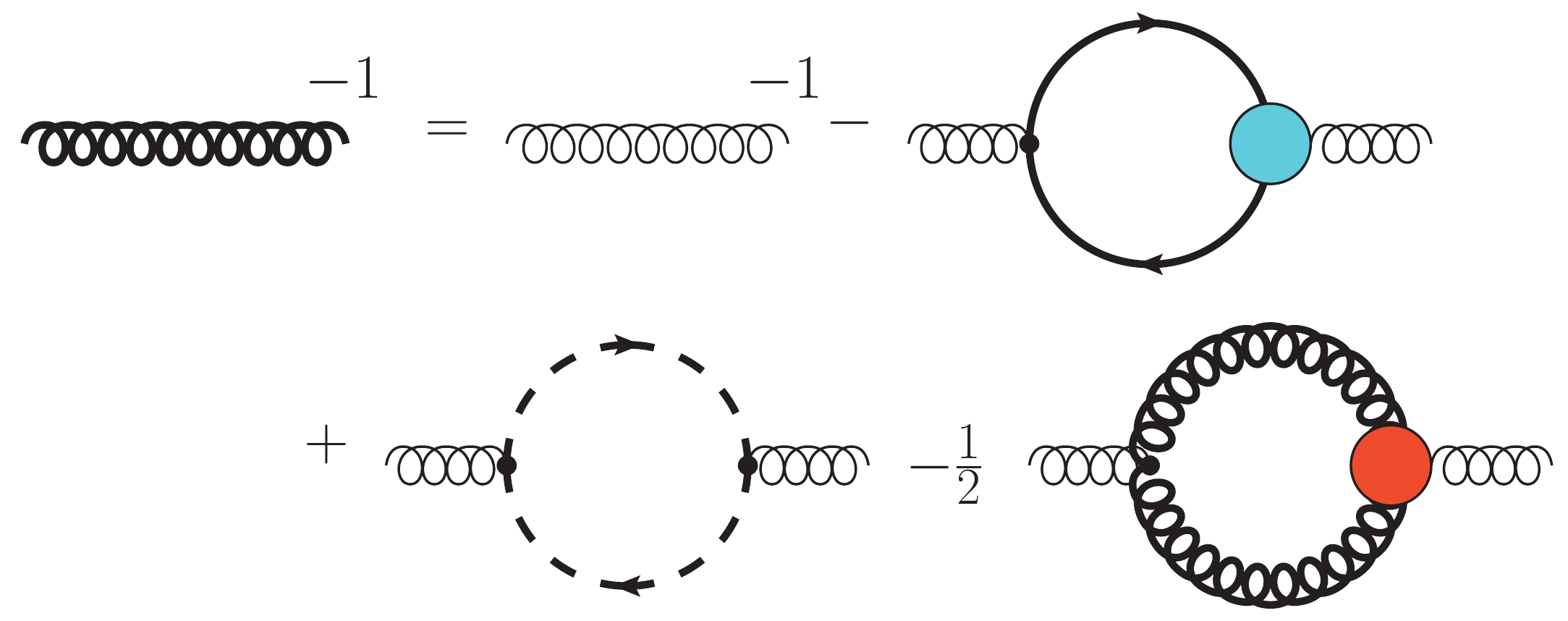}
\caption{The Yang-Mills system in a truncation corresponding to Ref.\
\cite{Fischer:2003rp} except that in our work a scalar loop has been employed. 
The ghost vertices are left bare.}
\label{Fig:CoupledSystemYM}
\end{figure}

Scalar QCD provides an excellent playground to
test, implement or improve numerical techniques and routines which can subsequently be applied 
in the fermionic system.
Despite its simplicity, however, there are some drawbacks as can be seen from
the DSE for the scalar propagator depicted in Fig.\ \ref{Fig:scalar_prop}. 
Compared to the fermionic case one obtains a much richer diagrammatic structure due to four-scalar and 
two-scalar-two-gluon vertices not present in QCD.\footnote{The constant  contribution from the tadpole terms  can
in principle be absorbed within the renormalization procedure  but nevertheless the remaining two-loop terms  very
likely contribute at least in the mid-momentum regime, {\it cf.}, 
Ref.\ \cite{Mader:2013ru}.} Additional diagrams also appear in the DSEs for the gluon
propagator and the scalar-gluon vertex, see Ref. \cite{Fister:2010yw} for details. But since our aim is to apply
the techniques to the fermionic case, a truncation which  meets the requirements of Ref. \cite{Hopfer:2013np} is
appropriate.  In Fig.~\ref{Fig:CoupledSystemScalar} and Fig.~\ref{Fig:CoupledSystemYM} the corresponding
matter sector as well as the Yang-Mills system is shown.
We note that in Ref. \cite{Hopfer:2013np} a DSE-like equation for the quark-gluon vertex has been employed,
adopted from the nPI-guided ansatz proposed in Ref. \cite{Alkofer:2008tt}.
Thus, all vertices in the scalar-gluon vertex DSE are dressed.\footnote{For 
details we refer the reader to Refs. \cite{Alkofer:2008tt,Berges:2004pu}.}
One gets a closed system if the three-gluon vertex is specified for both systems.
Here the model proposed in Ref. \cite{Fischer:2003rp} has been employed.
The last two diagrams on the right-hand side of the vertex equation are 
referred to as the \textit{Abelian} and the \textit{non-Abelian} diagram.
The colour traces yield prefactors of $-{1}/{2N_c}$ and
${N_c}/{2}$, respectively. We will show that this natural colour
suppression of the Abelian diagram by $N_c^2$ is additionally amplified dynamically.
 Indeed, in our setup depicted in Fig.\ \ref{Fig:CoupledSystemScalar} it is mostly driven
by the dynamics of the system as it will be shown that this suppression is also apparent
for adjoint scalars, i.e. if both diagrams own the same colour trace.

To dress the vertex we use the decomposition
\begin{equation}
 \varGamma^\nu(p,q;k) = \tilde A\, p^\nu + \tilde B\, q^\nu 
                      = ( \tilde A + \tilde B )\, q^\nu + \tilde A\, k^\nu
                      = ( \tilde A + \tilde B )\, p^\nu - \tilde B\, k^\nu,
\end{equation}
where $p^\nu$ and $q^\nu$ are the in- and outgoing momenta and $k^\nu=p^\nu-q^\nu$ is their
relative momentum. The dressing functions $\tilde A$ and $\tilde B$ depend on the three
invariants $p^2$, $q^2$ and $p\cdot q$.
Using the transversal projectors $P^{\mu\nu}(k)$ appearing in the equations one obtains
\begin{equation}
\label{Eq:A}
 P^{\mu\nu}(k)\varGamma^\nu(p,q;k) = A\, P^{\mu\nu}(k)\,q^\nu = A\, P^{\mu\nu}(k)\,p^\nu,
\end{equation}
{\it i.e.}, a single dressing function $A(p^2,q^2,p\cdot q)$ is enough in the transverse Landau gauge 
to include the vertex. 
The abbreviations $x\equiv p^2$, $y\equiv q^2$ and $\zeta\equiv {p\cdot q}/{|p||q|}$ will be used
in the following.

\subsection{Numerical treatment}
The coupled system depicted in Fig.\ \ref{Fig:CoupledSystemScalar} and Fig.\ \ref{Fig:CoupledSystemYM}
has been solved using standard techniques, where the calculations have been performed on 
Graphics Processing Units (GPUs). For the Yang-Mills part the numerical routines proposed
in Ref. \cite{Hopfer:2012ht} have been adopted as well as improved in order to allow for an arbitrary
number of GPU devices. For the matter sector a conventional integration grid has been employed 
using a non-linear
mapping for the Gauss-Legendre nodes. Furthermore, an internal integration grid
for the radial integrals is used
 which differs from the external one in order to avoid numerical instabilities.
A further advantage of this step is the numerical performance gain since it allows to calculate 
the vertex on a rather coarse external grid, where subsequently a cubic spline interpolation 
routine interpolates smoothly between the calculated external grid points and generates the values
needed for the finer internal grid. This step is performed within each iteration and
 can be done on GPUs in a very efficient way due to the high degree of parallelism. 
With this simple method one can reduce the amount of computing time without loss of accuracy 
resulting in running times
of only a few minutes for the whole system. 
Within the Abelian diagram further interpolation steps have to be done. Here a bi/tri-linear 
interpolation routine between the spline values has been implemented. While for the scalar case this seems to
over-shoot the problem, our aim is to adopt the methods to the fermionic case where it is
possible that the Abelian diagram behaves non-trivially. 
Furthermore, the renormalization procedure of the vertex system has been adopted from Ref.
\cite{Hopfer:2013np}.
\section{Results for the vertex dressing function}
The system described in Sec.\ \ref{Sec:scalarQCD} has been solved self-consistently including both
the Abelian and the non-Abelian diagram. In the following
we present our results for the vertex dressing function $A$ introduced in Eq.\ \eqref{Eq:A}.
The calculations have been performed
for scalars both in the fundamental as well as in the adjoint representation. 
\begin{figure}[h]
\centering
\subfigure{\includegraphics[width=6.33cm]{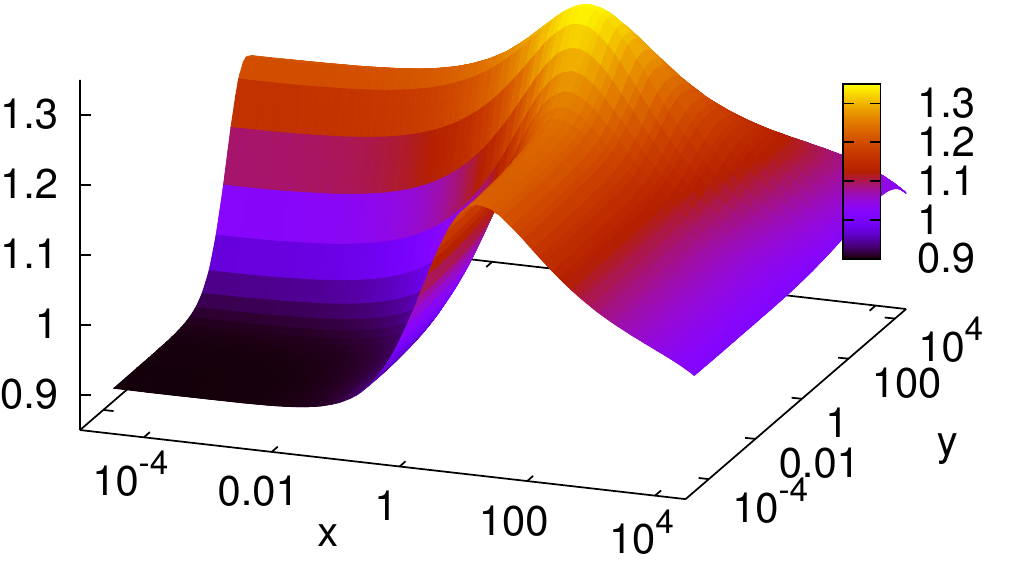}\label{Fig:VertexA}}
\hspace{-0.3cm}
\subfigure{\includegraphics[width=6.33cm]{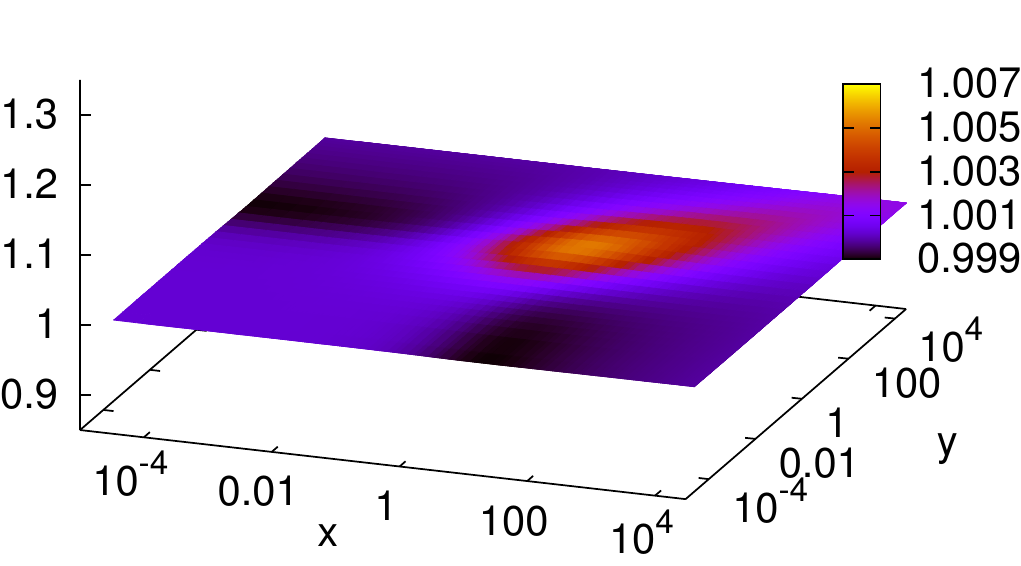}\label{Fig:VertexB}}
\subfigure{\includegraphics[width=6.33cm]{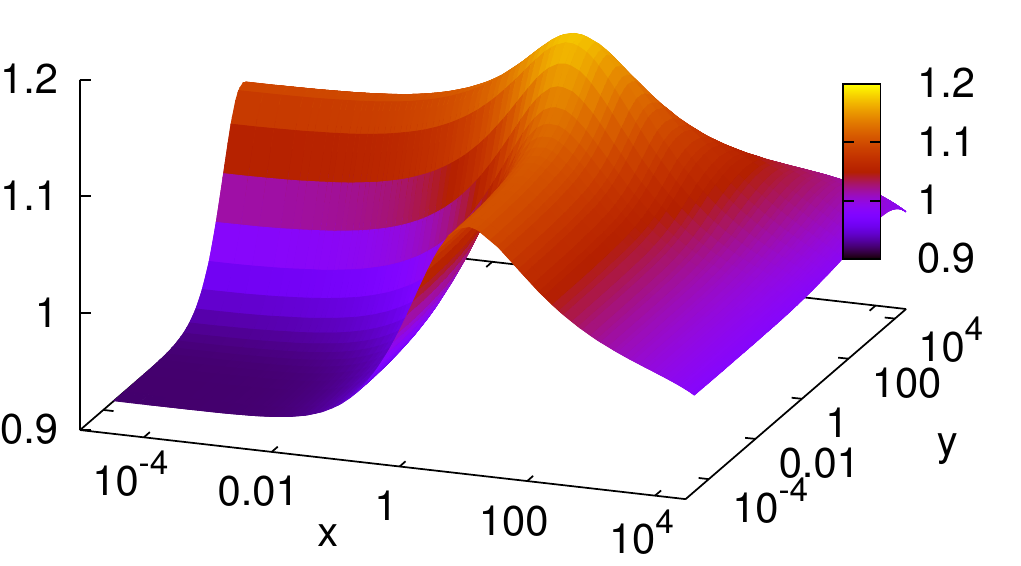}\label{Fig:Vertex_adj_A}}
\hspace{-0.3cm}
\subfigure{\includegraphics[width=6.33cm]{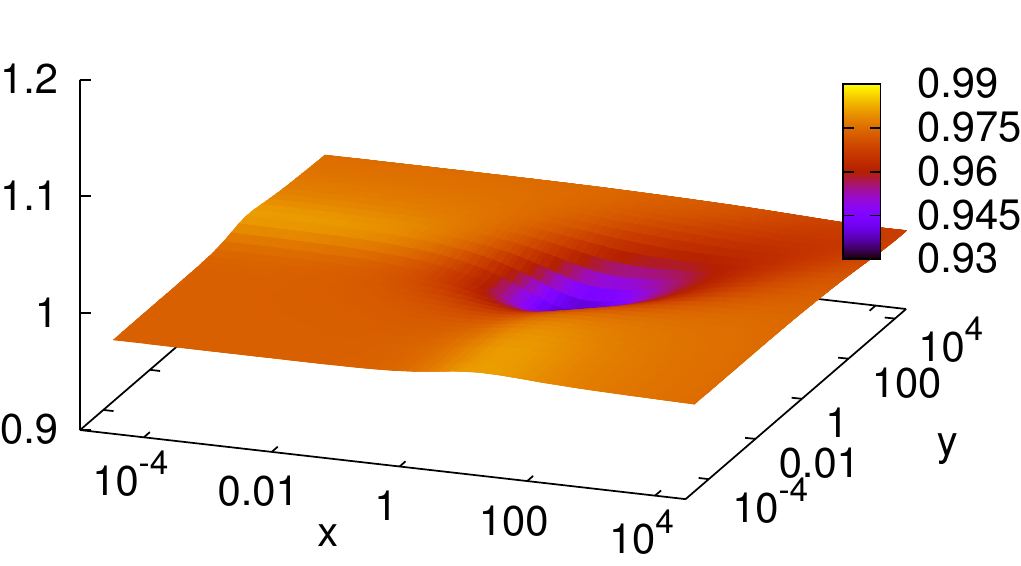}\label{Fig:Vertex_adj_B}}
\caption{
The vertex dressing function $A(x,y,\zeta)$ plotted for $\zeta\approx -1$.
The upper panel shows results for $SU(3)$ with a scalar field in the fundamental representation.
Non-Abelian contributions (left panel) are dominant. The results for the
adjoint representation of $SU(2)$ (lower panel)
reveal that the colour structure has a minor influence.
We also checked that this behaviour is hardly affected when varying the scalar masses.
Here, no feedback from the scalar loop has been taken into account.}
\end{figure} 
Fig.\ \ref{Fig:VertexA} shows the contribution of the non-Abelian diagram for
the fundamental representation. 
As can be seen from Fig.\ \ref{Fig:VertexB} the
Abelian diagram contributes only very little. Besides its natural colour suppression it is
 also suppressed dynamically. 
This is evident when one puts the scalar field into the adjoint representation.
Here the prefactors of the diagrams become the same, 
{\it i.e.}, the colour trace for the Abelian diagram changes from $-{1}/{2N_c}$ to ${N_c}/{2}$.
However, as can be seen from Fig.\ \ref{Fig:Vertex_adj_A} and Fig.\ \ref{Fig:Vertex_adj_B} also in this
case the Abelian diagram is suppressed and contributes only marginally.
We note that unquenching effects
result in a global suppression of both diagrams. Here, the non-Abelian diagram is more affected
due to its a priori larger contribution. 
Furthermore, in the adjoint representation the colour factor for the 
scalar loop changes from $\frac{1}{2}$ to $N_c$, and
the system is more sensitive to flavour changes. 
A more detailed investigation of all these and related effects is deferred to an upcoming study.

\section{Conclusions and outlook}
Scalar QCD in Landau gauge has been investigated using a Dyson-Schwinger/nPI-Action approach. 
The coupled equations for the scalar and Yang-Mills propagators as well as the scalar-gluon vertex 
have been solved self-consistently in a truncation adapted
to the fermionic system treated in Ref.\ \cite{Hopfer:2013np}. 
The calculations have been performed for 
scalar fields in both, fundamental and adjoint, representations. 
A strong dynamical suppression of the Abelian diagram
in the scalar-gluon vertex DSE has been found in both cases.
Therefore this effect is dynamical and 
independent of the colour structure to a high degree.
The numerical techniques can be transferred to the more complicated fermionic case \cite{Hopfer:2013np} 
in a direct way. 

\section*{Acknowledgments}
We thank Mario Mitter for helpful discussions. 
MH acknowledges support  from the Doktoratskolleg ''Hadrons in Vacuum, Nuclei and Stars`` of the 
Austrian Science Fund, FWF DK W1203-N16.  

\end{document}